\documentclass[showpacs,preprintnumbers,amsmath,amssymb]{revtex4}

\usepackage{graphicx}
\usepackage{dcolumn}
\usepackage{bm}
\begin{document}
\draft \font\Bbb =msbm10  scaled \magstephalf
\def\id{{\hbox{\Bbb I}}}

\title{Unconditional fidelity thresholds in
single copy distillation \\ and some aspects of quantum error
correction}

\author{
Pawe\l{} Horodecki}\email{pawel@mif.pg.gda.pl}
\author{Maciej Demianowicz}\email{maciej@mif.pg.gda.pl}
\address{Faculty of Applied Physics and Mathematics\\
Gda\'nsk University of Technology, 80--952 Gda\'nsk, Poland}

\begin{abstract}
Various aspects of distillation of noisy entanglement and some
associated effects in quantum error correction are considered. In
particular we prove that if only one--way classical communication
(from Alice to Bob) is allowed and the shared $d \otimes d$ state
is not pure then there is a threshold for optimal entanglement
fraction $F$ of the state (being an overlap between the shared
state and symmetric maximally entangled state) which can be
obtained in single copy distillation process. This implies that to
get (probabilistically) arbitrary good conclusive teleportation
via mixed state at least one classical bit of backward
communication (for Bob to Alice) has to be sent.
We provide several other threshold properties in this context
including in particular the existence of ultimate threshold of
optimal $F$ for states of full rank.

Finally the threshold results are linked to those of error
correction. Namely it is pointed out that in quantum computer
working on fixed number of quantum bits almost any kind of noise
can be (probabilistically) corrected only to some threshold error
bar though there are some (rare) exceptions.

\end{abstract}

 \maketitle

\newtheorem{definition}{Definition}
\newtheorem{theorem}{Theorem}
\newtheorem{lemma}{Lemma}
\newtheorem{conclusion}{Corollary}
\newtheorem{proposition}{Proposition}
\newtheorem{observation}{Observation}

\section{Introduction}
Quantum teleportation is one of the most interesting phenomena in
quantum information theory \cite{Be93}. In its original version it
involves perfect transmission of unknown spin $s$ state $\phi$
from Alice to Bob by means of shared maximally entangled pair
(called quantum channel) of two spin $s$ particles and $2s+1$
classical bits sent from Alice to Bob via some classical channel.
First step of the scheme is due to Alice. It involves the joint
complete von Neumann measurement of observable with maximally
entangled vectors Alice performs on both particle $\phi$ and one
member of maximally entangled pair. After getting the result Alice
sends it to Bob who performs some unitary operation on the second
member of the pair reconstructing the state $\phi$ though, in
general, neither Alice nor Bob know the state $\phi$ which is
teleported. However it is known that more general schemes
involving mixed states \cite{Po94}, conclusive teleportation
\cite{Tal} or general so-called LOCC operations \cite{gen} are
known. In particular it has been shown \cite{gen} that
optimization of teleportation fidelity $f$ under local operation
and classical communications is in one to one correspondence with
optimization of the so called singlet fraction $F$.

On the other hand there was a natural question how big $F$ Alice
and Bob can achieve given a single copy of strictly mixed (i.e.
not pure) quantum  state. It has been shown that for some highly
mixed states there exists an unconditional threshold value
$F_{threshold}<1$ (see \cite{Pop,Kent}). However it turned out
that there are some strictly mixed states for which Alice and Bob
can get as good $F$ as they want if they use general LOCC with
two--way classical communication in a conclusive way. In fact they
can achieve $F$ arbitrary close to unity but with the probability
$p(F)$ of achieving that going to zero with $F$ approaching unity.
This type of conclusive success called {\it quasidistillation} may
be important in some special cases of `game' like scenarios. This
leads to a surprising result that sometimes we can achieve
arbitrary good teleportation via mixed states (with the
probability of the process depending on the fidelity we require).

Now question is: is it possible to do the trick better, namely
involving only one--way classical communication from Alice to Bob
? In the present paper we show that it is impossible - in this
case the threshold value of optimal $F$ for any mixed state is
unconditional.

It follows in particular that there is a teleportation threshold
on transmission fidelity $f$ of unknown state from Alice to Bob
i.e. $f\leq f_{threshold}<1$. It must be stressed that this was
not obvious because, as it was recalled above, it is not true for
some mixed states if two--way communication is allowed. Further we
show that such threshold result is true in general (i.e. also if
two--way communication is allowed) when the state shared by Alice
and Bob is either of full rank or it is supposed to subject only
to tracepreserving LOCC operations.

\section{Improving singlet fraction of single copy -
general concepts and methods}

\subsection{Achieving maximal entanglement}
\label{Improvingmax}
Here the aim of the process is to achieve the maximal entanglement
which is admitted by the 'physics' of the system. Consider the
state described on the Hilbert space ${\cal H}$ of the total
system, ${\cal H}={\cal H}_{A} \otimes {\cal H}_{B}$, where
dimensions $d_{A}\equiv dim{\cal H}_{A}$, $d_{B} \equiv dim{\cal
H}_{B}$. Such systems are called in general {\it $d_{A} \otimes
d_{B}$ systems}.

For them the family of maximally entangled states has some
representative. This is the symmetric state
\begin{equation}
P_{+}=|\Psi_{+} \rangle \langle \Psi_{+} |, \ \ |\Psi_{+}
\rangle=\frac{1}{\sqrt{d}}\sum_{i=0}^{d-1} |i\rangle \otimes
|i\rangle, \ \ d=min[d_{A},d_{B}]. \label{max}
\end{equation}
Indeed any other maximally entangled state $P_{max}$ of
$d_{A}\otimes d_{B}$ system can be generated from (\ref{max}) by
the equation: $P_{max}=U_{1} \otimes U_{2}P_{+} U_{1}^{\dagger}
\otimes U_{2}^{\dagger}$ for some unitary operations $U_{1}$,
$U_{2}$.

The so-called singlet fraction \cite{gen} of the state $\varrho$
is defined as follows
\begin{equation}
F(\varrho)=\langle \Psi_+| \varrho |\Psi_+\rangle.
\end{equation}

Let us recall that Alice and Bob are allowed to perform {\it LOCC
operations}: the ones involving arbitrary local operations (LO) as
well as classical communication (CC). If given operation $\Lambda$
is LOCC we shall write $\Lambda \in$ LOCC. We have two classes of
LOCC operations:

(i) {\it unilocal} (or {\it one--way}) where only one party is
supposed to communicate the other, here we have two possibilities
$A \rightarrow B$ (Alice is allowed to call Bob) or $B \rightarrow
A$ (the opposite case),

(ii) {\it bilocal} (or {\it two--way}) if two parties can
communicate with each other.

For any of the above actions we have two possibilities. Namely any
$\Lambda \in$ LOCC can be

(i') {\it tracepreserving} i.e.  such that for any state $\varrho$
one has ${\rm Tr}(\Lambda(\varrho))=1$. Alice and Bob always have
to keep their particles after performing their actions,

(ii'') {\it conclusive} (see \cite{Tal}). Here  ${\rm
Tr}(\Lambda(\varrho))<1$. In this case Alice and Bob sometimes
throw away the particles if the result of their action is not
satisfactory.

We have also the following possibility of single copy
distillation. Suppose that Alice and Bob share only {\it one} copy
of $d \otimes d$ state. Then we have two ways of achieving maximal
entanglement:

(a) {\it single copy distillation} (SCD) under a chosen class of
operations ${\cal L} \subset $ LOCC takes place iff  Alice and Bob
can perform the action
\begin{equation}
\varrho \mathop{\longrightarrow}_{p{\mbox -probab.}}^{\Lambda \in
{\cal L}} P_+
\end{equation}
with probability of success $p={\rm Tr}(\Lambda(\varrho))>0$ (for
tracepreserving protocol we require $p=1$).

(b) {\it single copy quasidistillation} (SCQD) under a chosen
class of operations ${\cal L} \subset $ LOCC takes place if there
exists a sequence of $\Lambda_{n} \in {\cal L}$ such that
\begin{eqnarray}
\varrho \mathop{\longrightarrow}_{p_n{\mbox -probab.}}^{\Lambda
_{n} \in {\cal L}} \varrho_{n} \ {\mbox with} \ \ F(\varrho_n)
\rightarrow 1, \phantom{x} p_n \rightarrow 0 .
\end{eqnarray}

{\bf Remark .-} If $\varrho$ is SCD then it is also SCQD.

Let us recall that the operation is called {\it separable} iff it
is of the form
\begin{equation}
\varrho \rightarrow \varrho'\equiv\frac{\sum_{i} A_{i} \otimes
B_{i} \varrho A_{i}^{\dagger} \otimes B_{i} ^{\dagger} }{ {\rm
Tr}(\sum_{i} A_{i} \otimes B_{i} \varrho A_{i}^{\dagger} \otimes
B_{i} ^{\dagger}) } .
\end{equation}
In particular any LOCC superoperator is separable but not {\it
vice versa} \cite{nlwe}.

Some time ago the complete characterization of both SCD and SCQD
processes in two qubit case has been provided with help of the
Lorentz transformations technique \cite{Lorenz}. Below we shall
address general questions independent of the (finite) dimension of
involved Hilbert space.

\subsection{Achieving entanglement of smaller Schmidt rank}
The Schmidt rank (SR) of the bipartite (i.e. $d_{A} \otimes
d_{B}$) pure state is the rank of its reduced density matrices
(see \cite{Srank} for generalization of SR to mixed states domain)
equal to the number of its nonzero eigenvalues. In previous
section \ref{Improvingmax} we have consider methods of achieving
the state of maximal entanglement admitted by the physics of the
system. But given the $d_{A} \otimes d_{B}$ system there is a
possibility of more modest objective: to get pure state which is
maximally entangled under the constraint of SR bounded by some
$m\leq d=min[d_{A}, d_{B}]$. Family of such states has its
representative of the form
\begin{equation}
P_{+}^{m}=|\Psi_{+}^{m} \rangle \langle \Psi_{+}^{m} |, \ \
|\Psi_{+} ^{m} \rangle=\frac{1}{\sqrt{m}}\sum_{i=0}^{m-1}
|i\rangle \otimes |i\rangle \label{maxm}
\end{equation}
with the corresponding $m$-singlet fraction:
\begin{equation}
F^{m}(\varrho)=\langle \Psi_+^{m}| \varrho |\Psi_+^{m}\rangle.
\end{equation}
All the classification from the previous subsection apply i.e. we
have the corresponding so called {\it $m \otimes m$ SCD and SCQD
processes}. There is a remarkable result (see \cite{gen} ):
\begin{proposition}
A given $d_{A} \otimes d_{B}$ state $\varrho$ is $m \otimes m$ SCD
under separable operators iff there exists $m \otimes m$ product
projection $P \otimes Q$ such that $P \otimes Q \varrho P \otimes
Q$ is some pure (possibly unnormalized) projector of Schmidt rank
$m$. In particular for symmetric case  $d_{A}=d_{B}=d$ no mixed
state can be converted into the maximally entangled state of the
system. \label{mxm}
\end{proposition}

From the above we can prove immediately the following conclusion
\cite{Kent}:
\begin{conclusion}
The necessary condition for state $d_{A} \otimes d_{B}$ to be $m
\otimes m$ SCD is that the restriction on the rank:
$r(\varrho)\leq d_{A} d_{B}-m^{2}+1$.
\end{conclusion}

\section{Fidelity thresholds}
We define supremum of singlet fractions obtained from $\varrho$
under LOCC protocol belonging to the class $\{{ \cal C}\}$ as
$F_{sup}^{\cal C}(\varrho)$. We shall also single out ${\cal C}$
consisting of  $\rightarrow, \leftarrow, \leftrightarrow$ and
possibly with an additional index $x=d$ ($x=p$) standing for
deterministic (probabilistic). If we do not write $x$ it means
that we take  larger class  $x=p$. Note that the notion of
probabilistic protocol is equivalent to the notion of conditional
protocol.

 In analogy we define  $f_{sup}^{\cal C}(\varrho)$ as a teleportation fidelity
 that  is achieved under that class. It is known that for pure
 states probabilistic one way protocol gives
 $F_{sup}^{\cal \rightarrow}(\Psi)=\frac{r(\Psi)}{d}$ ($r$--Schmidt rank
 of $\Psi$) and is achieved in filtering protocol. In particular for
 $d \otimes d$ if $r(\Psi)=d$ then $F_{sup}^{\cal \rightarrow , \mathrm{p}}=1$
 can be achieved with finite probability.
The latter can be achieved by no mixed states. Still there are
states such that $F_{sup}^{\leftrightarrow}=1$ though this
supremum cannot be achieved with finite probability. Because of
the formula $f_{sup}^{\cal C}=\frac{F_{sup}^{\cal C}d+1}{d+1}$
that can be proven for all classes of protocols we get similar
conclusions for $ f_{sup}^{\cal C}$. Now we define

{\it Definition .- A singlet fraction threshold for $\varrho$ is
$F_{0}^{\cal C}=F_{sup}^{\cal C}$ iff the latter is strictly less
than one.}\\ In a similar way we define:

{\it Definition .- A teleportation fidelity threshold for
$\varrho$ is $f_{0}^{\cal C}=f_{sup}^{\cal C}$ iff the latter is
strictly less than one.}

We shall call thresholds proper if they are attained in some
protocol belonging to the required class ${\cal C}$. The following
formula is especially useful for seeking $F_{0}$:
\begin{equation}
F_{sup}^{\leftrightarrow}(\varrho)=
\mathop{sup}\limits_{||A||_{HS}=||B||_{HS}=1} F(\frac{A \otimes B
\varrho A^{\dagger} \otimes B^{\dagger}}{Tr(A \otimes B \varrho
A^{\dagger} \otimes B^{\dagger})})
\end{equation}

Note that one can take another trace--norm. One of the fundamental
observation is that for $d \otimes d $ states there are
teleportation thresholds that are not proper. An interesting
example of $F_{0}=\frac{2}{3}$ with such property is $2 \otimes 2
$ mixed state that can be quasidistilled (i.e. states from
\cite{TalPawel}) if embedded in $ 3 \otimes 3$ space. In general
we have an immediate conclusion:

{\it Observation .- Any SCQD $d\otimes d$ state that is embedded
into $d' \otimes d'$ space has the not proper fidelity treshold
$F_{0}=\frac{d}{d'}$.}

This follows easily from the fact (implied by convexity arguments)
that any $d\otimes d$ has a Schmidt rank at most $d$. After any
possible LOCC operation the rank cannot increase so the final
state will have the rank $d$ and can have at most fidelity
$\frac{d}{d'}$ with $d' \otimes d'$ singlet state $\Psi_{+}$.

\section{Thresholds in single copy distillation}
The simplest example when we have no threshold for $F$ is the
state \cite{gen}
\begin{equation}
\varrho=pP_+ + (1-p)|0\rangle\langle 0| \otimes |1\rangle \langle
1| \label{state}
\end{equation}
for $3 \otimes 3$ case.

Its counterpart for $2 \otimes 2$ case was introduced in
\cite{TalPawel}. The arbitrary good fidelity $F$ with nonzero
probability is achieved here by choosing $n$ high enough in the
following sequence of general quantum measurements $A_n,
\sqrt{I-A_n^2}$, $B_n,\sqrt{I-B_n^2}$ where
$A_n=diag[\frac{1}{n},1,1]$ and
$B_n=diag[1,\frac{1}{n},\frac{1}{n}]$ corresponding to Alice and
Bob actions. The conclusive result corresponding to the pair $A_n,
B_n$ turns $\varrho$ into $\varrho_n=\frac{1}{np+(1-p)}(npP_+ +
(1-p)|0\rangle\langle 0| \otimes |1\rangle\langle 1|) $ with
$F(\varrho_n)=\frac{np}{(n-1)p+1}$ approaching unity for large
$n$. The corresponding probability of the success however amounts
to $P(n)=\frac{1}{n}(p+\frac{(1-p)}{n})$. The latter goes to zero
with $n$ so we deal with SCQD effect here.

\subsection{Two--way communication case}
It is known that the trick with the state (\ref{state}) above does
not always work. Indeed the authors of Ref. \cite{Pop} have proven
that for so-called $2 \otimes 2$ Werner states there is no
possibility to increase $F$ at all. Moreover the same has been
proven for some entangled states perturbed by separable state of
high rank \cite{Kent}. Finally the low (four) rank $3 \otimes 3$
state has been shown to have threshold for $F$ too \cite{gen}. We
can prove the following:
\begin{observation}
Any $d \otimes d$ state of the form
\begin{equation}
\varrho=pP_{+} + \frac{(1-p)}{d} \sum_{i=1}^{d} |i\rangle \langle
i | \otimes |\pi(i) \rangle
 \langle \pi(i)|, \ p>0
\end{equation}
is not SCQD if $\pi$ represents arbitrary permutation of indices.
\end{observation}
{\it Proof .-} It is not difficult to see that the proof of Ref.
\cite{gen} (sect. VI, formula (39)) does not depend neither on the
dimension nor on the special character of chosen permutation of
the second basis in the separable state.

Here we have the limit on the SCQD process for a very special
class of states. The natural question is: what are the general
limits for SCQD trick like the one with state (\ref{state}) above?
In this section we shall prove some general sufficient conditions.

Following analysis of section
 provides the following \cite{gen}:

\begin{proposition}
 Any mixed $d \otimes d$ state is not SCD
under any separable (in particular LOCC) operation. In other words
it is not possible to turn single copy of strictly mixed  $d
\otimes d$ state into singlet state (\ref{max}) with finite
probability.
\end{proposition}

It is remarkable that the conclusion (\ref{mxm1}) above provides
suitable singularity of the state as necessary condition for SCD
but {\it not} for SCQD. Note, however that SCQD state
(\ref{state}) above still is singular.

Is it necessary for SCQD or, in other words,
  for  nonexistence of the threshold value for $F$ ?
It happens to be true as we have the following

\begin{proposition}
Any mixed $d_{A} \otimes d_{B}$ state of full rank is not $m
\otimes m$ SCQD under any separable operation. In particular there
exists threshold value of singlet fraction $F_{max}^{m}<1$ that
can be achieved with action of LOCC operations on $\varrho$.
\end{proposition}

The above property means that mixed $d_{A} \otimes d_{B}$ states
of full rank are never SCQD.

{\it Proof .-} It is easy to convince oneself (c.f.
\cite{Popescu}) that $\varrho$ is $m \otimes m$ SQCD under the
most general actions iff there exist a sequence of operators
$A_{n}$, $B_{n}$ such that
\begin{equation}
Tr(A_{n}^{\dagger}A_n)=Tr(B_{n}^{\dagger}B_n)=1 \label{norm}
\end{equation}
and the sequence
\begin{equation}
\varrho_{n}=\frac{A_{n} \otimes B_{n} \varrho A_{n}^{\dagger}
\otimes B_{n}^{\dagger} }{{\rm Tr} (A_{n} \otimes B_{n} \varrho
A_{n}^{\dagger} \otimes B_{n}^{\dagger})} \equiv \frac{A_{n}
\otimes B_{n} \varrho A_{n}^{\dagger} \otimes
B_{n}^{\dagger}}{c_n} \label{lim}
\end{equation}
satisfies
\begin{equation}
F_{m}(\varrho_{n})\rightarrow 1. \label{gran}
\end{equation}
Because $\varrho$ is of full rank we have $\varrho=p_{min}I\otimes
I + (1 - d^{2} p_{min}) \sigma $ for minimal eigenvalue
$p_{min}>0$ and some state $\sigma$. In this case the denominator
$c_n$ of (\ref{lim}) satisfies
\begin{equation}
c_{n}\geq p_{min}>0.
\end{equation}
Suppose now that (\ref{lim}) were true. Then according to
(\ref{norm}) we could choose subsequences of $A_{n}$, $B_{n}$
convergent to some $A$, $B$.

Now let us observe that the set of all $A_{n}$ satisfying ${\rm
Tr}(A_{n}^{\dagger}A_{n})=1$ can be interpreted as a sequence of
the points on a sphere with unit radius \footnote{It follows from
the fact that the set of all operators on finite--dimensional
space ${\cal C}^{d_{A}}$ (${\cal C}^{d_{A}}$) forms the
Hilbert-Schmidt space with the norm $||B|| \equiv \sqrt{{\rm
Tr}(B^{\dagger}B})$ induced by scalar product $\langle B,C \rangle
\equiv {\rm Tr}(B^{\dagger}C)$.} which is a compact set. Any
sequence formed by compact set elements has subsequence which has
the limit point. So in the quasidistillation process we can
restrict ourselves to such subsequence $\tilde{A}_{n}$ of $A_{n}$
with the limit operator $A$. The same holds for the sequence $B_n$
implying in an analogous way existence of the corresponding limit
operator $B$.

Now let us take $\alpha, \beta >0$ such that operators $I
-\alpha^2A^\dagger A$ and $I -\beta^2 B^\dagger B$ have positive
spectra. It is easy to see that then the result of the conclusive
bilocal action according to limit operators $A'=\alpha A$ and
$B'=\beta B$ would give the maximally entangled $m \otimes m$
 state $P_{+}^{m}$ with the finite probability $p\geq \alpha\beta p_{min}$.
This would mean that $\varrho$ were $m \otimes m$ SCD. Following
Observation \ref{mxm1} the latter is in the contradiction with
assumption that $\varrho$ has full rank. This concludes the proof.

{\bf Remark .-} One can ask whether the proposition could be
extended to have  ``if and only if'' form. It is impossible -
there are states (see \cite{Kent,gen}) which are not of full rank
but still have threshold value for $F$. The most surprising seem
to be examples of Ref. \cite{gen} which have relatively small rank
($4$ in the presence of the full rank $9$). However the proof of
existence of the threshold in that case is not straightforward at
all.

But we can modify the above proposition in other way making it
still true. Namely we can relax the condition about full rank of
the state imposing at the same time tracepreserving property on
LOCC operations:

\begin{proposition}
If Alice and Bob share any strictly mixed $d \otimes d$ then there
always exists threshold value of singlet fraction $F_{max}<1$
Alice and Bob can achieve with help of any tracepreserving LOCC
operations.
\end{proposition}

{\it Proof .-} The proof is straightforward. It is sufficient to
prove that the state is not SCQD. If it were then there would
exist the sequence of separable tracepreserving superoperators
$\Lambda_{n}$ mapping $\varrho$ into the sequence of states
providing $P_+$ in the limit. However the tracepreserving
superoperators (channels) correspond to a special class of states
\cite{gen,Cirac}. In particular under a special choice of
isomorphism \cite{Cirac} the separable superoperators correspond
to some separable states on higher Hilbert space. The set of those
separable states is compact. Now {\it via} the isomorphism if some
mixed state were SCQD then there would exist the limit separable
superoperator which would be tracepreserving. It is easy to see
that it would turn the state under consideration into $P_+$. But
then we would have the mixed state which would be SCD which is
impossible (see Proposition. 1). This concludes the proof.

\subsection{Proof of unconditional
'one--way' threshold for mixed states}
Consider the following:

{\bf Problem .- }{\it Alice and Bob are allowed to perform all
LOCC operations with one--way classical communication (either from
Alice to Bob or vice versa). Suppose that they share the state
which is not pure. Is it possible for any $\tilde{F}$ (chosen to
be however close to $1$) to transform the $\varrho$ into the state
$\varrho'$ with $F(\varrho')=\tilde{F}$ ?}

{\bf Remark .-} In the above we ask just about SCQD property of
the state (see sect. II).

We shall solve the above Problem (which is the main result of the
paper) negatively. It will be achieved by showing that if one--way
quasidistillation were possible (i.e. if the above problem had
positive solution) then unilocal SCD of some mixed state would be
possible which is, however, forbidden by general property 1.

%

Consider the case when Alice is allowed to phone  Bob but not vice
versa. Because the singlet fraction $F$ is convex in $\varrho$ it
is enough to consider only

unilocal ($\equiv$ one--way) operations in a very special
conclusive process.
In fact, as we shall see below the possibility of unilocal
conclusive distillation would be equivalent to possibility of
conclusive distillation via the operations
\begin{equation}
\varrho_{AB} \rightarrow \Lambda_{uniloc}(\varrho_{AB})=\frac{V
\otimes I (I \otimes \Lambda_{TP}(\varrho_{AB})) V^{\dagger}
\otimes I} {Tr(V \otimes I (I \otimes \Lambda_{TP}(\varrho_{AB}))
V^{\dagger} \otimes I)} \label{form}
\end{equation}
with some operator $V$,  $Tr(V^{\dagger} V)=1$, utilized by Alice
and some {\it tracepreserving} (TP) Bob's  POVM $\Lambda_{TP}$.
The possibility of the above restriction can be shown as follows.
The most general Alice action is {\it any} POVM with results $k=1,
..., K$. After getting any result Alice phones Bob who can perform
POVM depending on Alice result but without subselection i.e.
throwing away his particle in some cases. The latter process is
forbidden because Bob cannot inform Alice about his possible
decision not to keep the particle. So the decision would result in
the situation when Alice keeps her particle while Bob does not.

Thus Bob POVM-s {\it must} be tracepreserving. Note that if Alice
and Bob can achieve some fidelity $\tilde{F}$ via the above
procedure then they {\it must} achieve it after {\it one} of Alice
results. Otherwise any member of resulting mixture (due to
different Alice's results) would have $F < \tilde{F}$ and this
would follow from the convexity of the mixture. This would also
imply that the whole state would not reach fidelity $\tilde{F}$.
Summarizing: we have reduced the problem to {\it one} result of
Alice POVM. This leads to Alice action of type $\frac{ V \otimes I
(\cdot) V^{\dagger} \otimes I}{{\rm Tr}(V \otimes I (\cdot)
V^{\dagger} \otimes I)}$. This is nothing but usual filtering
operation (\cite{Gisin}, see also \cite{pur,xor} for powerful
application of that kind of operations).

As the last operation is invariant under multiplication of $V$ by
any number we can restrict ourselves to filters $V$ satisfying
${\rm Tr}(V^{\dagger}V)=1$ which leads to the form (\ref{form}) of
Alice and Bob actions.

As we promised before now we shall prove that nonexistence of
threshold for $F(\varrho_{AB}')$ with $\varrho_{AB}'$ resulting
from the unilocal transformation (\ref{form}) would imply single
copy distillability of $\varrho_{AB}$.

By the very definition, nonexistence of the threshold for $F$
implies that there exists some sequence of such actions
$\Lambda_{uniloc}^{n}$ of the form (\ref{form}) and some mixed
state $\varrho_{AB}$ such that
\begin{equation}
\Lambda_{uniloc}^{n}(\varrho_{AB}) \rightarrow |\Psi_{+} \rangle
\langle \Psi_{+}|. \label{limit}
\end{equation}
Then first it follows that {\it Alice reduced density matrix
$\varrho_{A}$ has to be of maximal rank}. In fact otherwise all
pure states in mixture would have strictly less Schmidt rank than
$|\Psi_{+}\rangle$. As no bilocal (hence in particular unilocal)
action can increase Schmidt rank \cite{Lo} we get immediately that
all pure states resulting in mixture would have always Schmidt
rank less than $N$ so they could not converge to maximally
entangled state of Schmidt rank $d$.

Now let us utilize the fact that (\ref{limit}) implies that the
reduced sequence of reduced density matrix describing Alice
particle satisfies
\begin{equation}
Tr_{B}(\Lambda_{uniloc}^{n}(\varrho_{AB})) \rightarrow \frac{I}{d}
\label{chaos}
\end{equation}
i.e. it converges to maximally chaotic state.

Bob's action -  as the tracepreseving one -  can have no impact on
the form of reduced density matrix describing Alice particle (this
is consequence of nonexistence of superluminal signaling within
quantum mechanics). In other words we have
$Tr_{B}(\Lambda_{uniloc}^{n}(\varrho_{AB}))= Tr_{B}(\frac{V_{n}
\otimes I \varrho_{AB} V_{n} \otimes I }{Tr(V_{n} \otimes I
\varrho _{AB} V_{n} \otimes I)})$ which means that Alice alone
must force the density matrix of her particle to satisfy
(\ref{chaos}). All this implies
\begin{equation}
\frac{V_{n} \varrho_{A} V_{n}}{{\rm Tr}(V_{n} \varrho_{A} V_{n})}
\rightarrow \frac{I}{d}
\end{equation}
(recall that $\varrho_{AB}$ is of $d \otimes d$ type). By virtue
of the previous analysis we know that the state $\varrho_{A}$ is
nonsingular. So its spectral decomposition allows for the
representation
\begin{equation}
\rho_{A}=p_{min}I + (1-dp_{min})\sigma_{A}
\end{equation}
for some singular state $\sigma_{A}$ and $p_{min}>0$ being minimal
eigenvalue of $\varrho_{A}$. Furthermore the denominators
constituting the normalization factor after Alice filtering
operations $V_{n}$ satisfy
\begin{eqnarray}
&&{\rm Tr}(V_{n} \varrho_{A} V_{n}^{\dagger})=
{\rm Tr}(V_{n}^{\dagger} V_{n} \varrho_{A} )= \nonumber \\
&&p_{min}Tr(V_{n}^{\dagger} V_{n}) +(1 - d p_{min})
Tr(V_{n}^{\dagger} V_{n}\varrho'_{AB})\geq d p_{min}>0
\end{eqnarray}
as we assumed $Tr(V_{n}^{\dagger}V_{n})=1$. Hence all the
denominators are {\it separated} from zero by the same constant $d
p_{min}$.

Again as in sect. III operators $V_{n}$ are normalized via
condition $Tr(V_{n}^\dagger V_n)=1$ thus they form a compact set
and some of their subsequence converges to some operator $V$.

The set of corresponding Bob POVM-s form the compact set too.
Indeed as tracepreserving ones they correspond to some sequence of
bipartite quantum states \cite{gen}. So again, their subsequence
$\tilde{\Lambda}_{TP}^{n}$ converges to some limit tracepreserving
POVM $\tilde{\Lambda}_{TP}$. Thus we get some limit Alice and Bob
action defined by $\tilde{V}$ and $\tilde{\Lambda}_{TP}$. This
limit action is well defined on the state $\varrho_{AB}$ as via
continuity argument the Alice normalizing factor is separated form
zero i.e. no singularity like for the action of type leading to
quasidistillation of states (\ref{state}) occurs.

Continuity of $F(\varrho_{AB})$ function implies that the limit
output state $\varrho_{AB}^{limit}$ achieved by Alice and Bob
satisfies
\begin{equation}
\varrho_{AB}^{limit} \equiv \frac{\tilde{V} \otimes I ( I \otimes
\tilde{\Lambda}_{TP}(\varrho_{AB})) \tilde{V}^{\dagger} \otimes I}
{Tr(\tilde{V} \otimes I (I \otimes \Lambda_{TP}(\varrho _{AB}))
\tilde{V}^{\dagger} \otimes I)} =|\Psi_{+} \rangle \langle
\Psi_{+}| \label{eq}.
\end{equation}
Thus we have shown that under one--way communication and one copy
restrictions the existence of quasidistillation implies
distillation in this case. But the latter is impossible for
strictly mixed state (see proposition 1).

Now we can conclude one of the  main results of the present paper:

\begin{proposition}
\label{one-wayTh} If Alice and Bob share bipartite $d \otimes d$
state which is strictly mixed
  and only one-way LOCC actions are allowed
(either from Alice to Bob or vice versa) then there is a singlet
fraction threshold $F_{thr}<1$.
\end{proposition}

{\bf Remark .-} If the shared state is not mixed the statement is
not true. It is simple to see that it is violated by any entangled
state of maximally $d$ Schmidt rank. Similarly if Alice spin is
strictly greater than Bob one than sometimes it is possible to get
perfect distillation. The example of such process is given in
\cite{gen}. Finally if Bob has his spin greater than Alice it is
not difficult to see that still one can arrange the situations
when the members of the mixture are locally orthogonal (see
\cite{term}) on Bob side. Then for some states, even without Alice
actions,  Bob can turn $F$ into unity on his own.

\section{Teleportation threshold}

Now  we shall briefly show that there is teleportation threshold
via mixed states even if the {\it conclusive or conditional}
teleportation is allowed. In Ref. \cite{gen} it has been shown
that optimal teleportation fidelity $f_{max}$ using  full class of
LOCC operations satisfies
\begin{equation}
f_{max}=\frac{dF_{max}+1}{d+1} \label{formula},
\end{equation}
where $F_{max}$ is maximal singlet fraction obtained by general
(bilocal conclusive) LOCC actions. The proof of the above formula
does not change if instead of general LOCC class we require
another subclass of LOCC operations from four possible ones (see
sec. II): bilocal tracepreserving, bilocal conclusive, unilocal
tracepreserving, unilocal conclusive. It is implied by the fact
that operation belonging to any of these classes does not change
its class if followed by so-called $U \otimes U^*$ twirling
operation (see \cite{gen}). This is because twirling is a
tracepreserving operation which can be implemented with one-way
classical communication (from Alice to Bob or vice versa, it does
not matter).

The formula (\ref{formula}) immediately implies that all our
results (propositions 2, 3, 4) providing thresholds for singlet
fraction $F_{thr}<1$ imply also thresholds for teleportation
\begin{equation}
f_{max}=\frac{dF_{max}+1}{d+1}\leq\frac{dF_{th}+1}{d+1}\equiv
f_{th} < 1.
\end{equation}
In particular if Alice and Bob are allowed to form general
unilocal channel (conclusive or not) then the above threshold
always occurs.

\section{Quantum states and quantum channels: some interesting aspects}
In general {\it quantum channel} \cite{huge} $\Lambda$ is a
tracepreserving, completely positive map (superoperator) with its
usual origin in environment. There is the notion of {\it quantum
channel capacity} $Q_C=Q_{C}(\Lambda)$ \cite{huge}. This is the
maximal possible rate $k/n$ such that $k$ qubits states (or their
equivalents embedded in higher Hilbert space) can be reliably sent
to Bob down the new channel $\Lambda^{\otimes n}$ (composed of $n$
copies of the original channel) with help of classical resource
$C$. Here $C=\leftrightarrow,\leftarrow,\rightarrow,\o$
corresponds to two--way, one--way backward, one--way forward and
zero--way classical communication from Alice to Bob. One of the
basic result is that $Q_{0}=Q_{\rightarrow}$ \cite{Howard,huge}
while any protocol corresponding to $Q_{0}$ represents some  error
correcting code. Similarly one defines distillable entanglement
$D_{C}(\varrho)$ as the maximal possible rate $k/n$ such that $k$
singlets one can get from $\varrho^{\otimes n}$ in the limit of
large $n$. There is a basic inequality
$D_{C}(\varrho(\Lambda))\leq Q_{C}(\Lambda)$ \cite{huge}.

There is also a connection between quantum channels and quantum
states \cite{Schumacher}:
\begin{equation}
\varrho_{\Lambda}=[\id \otimes \Lambda](P_{+}) \label{stchan}
\end{equation}
which was fruitfully applied in quantum data compression
\cite{Schumacher,Howard} and for example in binding entanglement
channels construction.

{\bf Remark (Counterfactual error correction).-} Any process of
reliable transmission of information through noisy quantum channel
$\Lambda$ is called {\it quantum error correction}
\cite{Springer}. It is known that $Q_{\empty}=Q_{\rightarrow}$
i.e. typical error correction consisting of Alice encoding,
transmission and Bob decoding is as good as apparently more
'powerful' process supplemented by one--way classical information
channel (from Alice to Bob). However, in general
$Q_{\leftrightarrow}>Q_{\rightarrow}$. The process with resource
$C=\leftrightarrow$ comprises some unusual error correction which
has no classical counterpart. This relies on sending number of
entangled pairs down the channel, entanglement distillation
procedure and final teleportation \cite{huge}. This can be
interpreted as {\it counterfactual error correction}
\cite{Springer}. Summarizing: in quantum data transmission through
channel $\Theta$ instead of usual error correction achieving
capacity $Q_{\rightarrow}(\Theta)=Q_{\empty}(\Theta)$ one can
produce the state $\varrho_{\Theta}$ and achieve better
transmission capacity distilling first
$D_{\leftrightarrow}(\varrho_{\Theta})$ singlets an then
teleporting the information. In seminal paper \cite{CJEP} it has
been shown that $D_{C}(\varrho_{\Theta}) \leq Q_{C}(\Theta)$
($C=\leftarrow,\leftrightarrow$) which implies that nonzero $D$
implies nonzero $D$ in those cases. Quite remarkably, the converse
also happens to be true \cite{CJEP}. This allows to answer some
questions about perfect error correction. The general answer has
been already obtained \cite{Raginsky}, however here we shall
reproduce he result basing mainly on entanglement distillation
concept.
\section{Probabilistic error correction and its limits}
We shall be interested in the {\it probabilistic} or {\it
conclusive} error correction. Suppose that Alice sends some
(encoded) quantum information down the channel which is decoded by
Bob afterwards. Finally they want that (i) Bob gets the perfectly
correct information with some nonzero probability $p$ (ii) they
both know when this happens. Similarly Alice alone can be
interested in running quantum computer program. She wants to get
perfect result, but she allows it to happen probabilistically.
This is especially important where possible wrong result could
spoil more than good one could help her future decision. This idea
of probabilistic error correction is parallel to the one of
probabilistic algorithms.

In what follows we shall assume that the single quantum bits
(qubits) or, in general n-bits will subject to interaction with
environment independently i.e. we assume uncorrelated errors (this
is common assumption in quantum computing theory \cite{Zurek}).

Now the first basic question is:

{\it Question A .- When the perfect probabilistic error correction
is possible ?}

Here by perfect correction we mean that we just send {\it all the
information} without coding \footnote{One may also allow for
unitary coding here without changing the subsequent conclusions.}
down the channel and then try to correct errors that happen to the
already sent information.

However sometimes we could be interested in quasi-probabilistic
error correction i.e. when we can correct better and better but
with the probability decreasing to zero \footnote{If it does not
decrease then because of finite dimension of quantum computer
always exists the limit error correcting scheme that corresponds
to probabilistic error correction considered before.}.

{\it Question B .- When the quasi-probabilistic error correction
is possible ?}

The detailed answer is not easy but we shall give some definite
answer:

{\it Proposition .- Under any kind of errors the channels allowing
for perfect probabilistic error correction are of zero measure in
the set of all channels. However there are some channels, among
them some satisfying local error assumption, that allow for either
perfect probabilistic or quasi-probabilistic error correction.}

Proof .- We exploit the isomorphism between quantum states and
quantum channels (see for instance \cite{gen}). Any quantum
channel $\Theta: {\cal H}_{A} \rightarrow {\cal H}_{B}$ is
isomorphic to some quantum bipartite state $\varrho_{AB}$ with
maximally mixed subsystem $A$. The natural measure on quantum
channels is then a measure on this class of states. If there is no
restrictions then the measure is just a product of simplex measure
(corresponding to spectrum) and Haar measure (representing
eigenvectors) \cite{volume}. For states corresponding to channels
the measure is more complicated since the condition on
$\varrho_{A}$ must be taken into account. Still it is easy to see
that any subset of channels that corresponds to singular states is
of measure zero in this picture.

It easily follows from the theorem \ref{one-wayTh} that it is
still impossible to correct probabilistically all errors since
then instead of sending the information we might send half of the
maximally entangled state down the channel, then achieve
probabilistically maximal fidelity in one-way scenario. This
concludes the proof.

Finally let us provide the channels that allow for
quasi-probabilistic error correction. Consider the state
\begin{equation}
\varrho_{AB}= \frac{b}{a+b} |\Psi\rangle\langle \Psi| +
\frac{a}{a+b} |0\rangle\langle1|, \ |\Psi\rangle = a|00\rangle +
b|11\rangle, \ \ a,b > 0.
\end{equation}
It is relatively easy to see that this state is quasidistillable
as the original one from \cite{TalPawel}. This means that in
one--way protocol one can achieve arbitrary high fidelity $F$. But
this also means that one can achieve (via sending half of the
singlet, one--way LOCC probabilistic action and final
teleportation) arbitrary good transfer of quantum data.

\section{Achieving fidelity threshold for one--way filtering
of maximally entangled mixed states}
Let us consider a state
\begin{equation}
\rho =  p |\Psi _{+} \rangle \langle \Psi _{+}|
 + (1-p) |01 \rangle \langle 01|.
\end{equation}
The main goal of this section is to find
\begin{equation}
\label{func} {\mathcal{F}} = \mathop{\max}\limits_{A} \; F \left(
\frac{ \mathrm I \otimes A \rho I \otimes A}{ \mathrm{Tr}[ I
\otimes A \rho I \otimes A ]} \right)
\end{equation}
with  $F$ being a singlet fraction.
 Maximization runs over all
hermitian operators of the form
\begin{equation}
\mathrm A=\alpha I + \vec{k} \cdot \vec{\sigma},
\end{equation}
where
\begin{equation}
\vec{k} = (k_{1},k_{2},k_{3}), \;\;\; \vec{\sigma} = (\sigma _{x},
\sigma _{y},\sigma _{z}),
\end{equation}
such that $\mathrm ||{A}||_{\mathrm{Tr}} \equiv
\mathrm{Tr}[|{A}|]=1$. Trace norm condition gives
\begin{equation}
|\alpha + k |+|\alpha - k| = 1
\end{equation}
and divides the range of parameters into three subranges. It is
known \cite{pur} that
\begin{equation}
\label{nodro} {\mathcal{F}}=\frac{1}{4}(1+{\mathrm{N}}(\rho)),
\end{equation}
where $\mathrm{N}(\rho)=Tr \sqrt{T^{\dagger} T}$. Function
$\mathrm{N}$ its maximum equal to $\frac{5}{3}$ reaches on the
boundaries. This value corresponds to $\alpha = \pm \frac{1}{2}$,
$\vec{k}=(0,0,\pm \frac{1}{4})$. From (\ref{nodro}) we then get
${\mathcal{F}}=\frac{2}{3}$. The operation Bob applies to his
subsystem is then of the form
\begin{equation}
A= \left(
\begin{array}{cc}
\pm\frac{3}{4} & 0\\
0 & \pm\frac{1}{4}
\end{array}
\right).
\end{equation}
\newline
Considerations above can be generalized to the case when filtering
is $\Lambda _{TP} \otimes \Lambda _{A}$.
\section{Summary and discussion}

Let us summarize the result of the paper. We have considered the
problem of resource of single copy of entangled state. Following
previous approaches the state is called single copy distillable
(SCD) if it can be converted (with provided additional resources)
with finite probability into maximally entangled state. It is
called single copy quasidistillable (SCQD) if it can be converted
with nonzero probability into the state arbitrarily close to
maximally entangled state but with the probability going to zero
when the output state approaches the maximally entangled state. If
some state $\varrho$ is not SCQD under given resources it means
that there is a threshold value for singlet fraction
$F(\varrho)\leq F_{th} <1$ Alice and Bob can achieve with help of
those resources. We have pointed out that state $d \otimes d$
state which is truly mixed (not pure) is not SCD. We have shown
that it is also not SCQD if only tracepreserving separable
superoperators are allowed. When the operations are not
tracepresering it was known before that some of the states are
SCQD and some others not. We have shown that among the latter
there are all states of full rank. Finally we have shown that if
Alice and Bob are allowed to use only one--way classical
communication then any truly mixed state is not SCQD i.e. in this
case we have unconditional threshold value for $F$. We have
pointed out that all cases where states are not SCQD (threshold
for $F$ occurs) there is also threshold for the teleportation
fidelity value $f_{th}<1$. Thus in order to get arbitrarily good
teleportation from Alice to Bob via truly mixed state {\it at
least one classical bit of backward communication (from Bob to
Alice) must be sent}.

Finally note that in Ref. \cite{aktyw} it has been shown that
there are states which are not SCQD but can allow
quasidisillability with help of arbitrary amount of bound
entangled (i.e. entanglement which can not be distilled) states
and two--way LOCC actions. This was the first effect showing that
bound entanglement  is somehow useful (see
\cite{super,concentration} for multiparticle effects of similar
type). It is interesting to ask whether the result of \cite{aktyw}
can be made stronger by allowing one--way LOCC action and bound
entanglement supply. In particular, as (according to the present
results) {\it any} mixed state has a one--way LOCC threshold value
for $F$ there is a more specific question: whether for {\it any}
mixed state there exists such form of bound entanglement that if
considered as a free resource can allow for quasidistillation of
the state in one--way LOCC regime.

The work was supported by European Union under grant RESQ No.
IST-2001-37559 and grant QUPRODIS No. IST-2001-38878, and by
Polish Ministry of Science under grant No. PB2-MIN-008/P03/2003.


\begin{references}
\bibitem{production}
This can be done as follows. Suppose that Alice and Bob have a
$d_{B} \otimes d_{B}$ state $\varrho$ and want to produce
$\varrho'=(1-\alpha)\frac{I}{d}+ \alpha \varrho^{\otimes n}$,
$d=d_{A}d_{B}$. They can use the following scenario: take the
system in a given state $\varrho$ and with probability $1-\alpha$
do nothing while with probability  $\alpha$ make the local `swap'
(or `flip') interaction $V$ between their local subsystems and
local ancillas prepared in maximally chaotic states $\frac{I}{d}$.
\end{references}
\end{document}